\documentclass[fleqn,10pt]{wlscirep}
\usepackage[utf8]{inputenc}
\usepackage[T1]{fontenc}
\usepackage{float}

\title{Artificial neural networks for density-functional optimizations in fermionic systems}

\author[1]{Caio A. Custódio}
\author[1]{Érica R. Filletti}
\author[1*]{Vivian V. França}

\affil[1]{Institute of Chemistry, S\~{a}o Paulo State University, UNESP, 14800-090, Araraquara, São Paulo, Brazil}

\affil[*]{vivian.franca@unesp.br}

\begin{abstract}
  In this work we propose an artificial neural network {functional} to the ground-state energy of fermionic interacting particles in homogeneous chains described by the Hubbard model. Our {neural network functional was proven to} has an excellent performance: it deviates from numerically exact calculations by less than $0.15\%$ for all the regimes of filling factors and magnetizations, and for a vast regime of interactions. When compared to analytical functionals, {the neural functional was found} to be more precise for all the regimes of parameters, being particularly superior at the weakly interacting regime: where the analytical parametrization fails the most, $\sim7\%$, against only $\sim0.1\%$ for the neural {functional. We have also applied our homogeneous functional to finite, localized impurities and harmonically confined systems within density-functional theory (DFT) methods. The results show that while our artificial neural network approach is substantially more accurate than other equivalently simple and fast DFT treatments, it has similar performance than more costly DFT calculations and other independent many-body calculations, at a fraction of the computational cost.}
   
\end{abstract}

\begin{document}

\flushbottom
\maketitle

\thispagestyle{empty}

\section*{Introduction}

Theoretical approaches in the quantum mechanical level are essential to the full understanding of the matter and its properties. Although the wavefunction of a given system $-$ and thus any property of that system $-$ may in principle be obtained by solving the Schrödinger equations, this exact method is extremely costly for many-particle interacting systems, despite the availability of high computational resources. This difficulty is enhanced further when the theoretical simulations incorporate also spatial inhomogeneities $-$ such as borders, impurities, {confining} potentials and disorder $-$ which are sometimes desired \cite{tobi} but in most of the cases {\it unavoidable} in realistic systems and experimental setups.

In this context the density-functional theory (DFT) \cite{PhysRev136B864,Kohn1999,Capelle2006} appears as an interesting and powerful tool. Within DFT the particle density of a system with $N$ interacting particles is used as the central quantity, instead of the wavefunction. Thus the task of obtaining $\Psi(\textbf{r}_1, \textbf{r}_2,...,\textbf{r}_N)$ $-$ a $3N$-dimensional function $-$ is reduced to obtain $ n $\textbf{(r)} $-$ a 3-dimensional function. This illustrates how much simpler DFT calculations are in comparison to wavefunction-based quantum calculations. The one-to-one mapping between $n$ and $\Psi$ is guaranteed by the Hohenberg-Kohn (HK) theorem \cite{PhysRev136B864}, which has been recently formally proved to be valid also for lattice systems \cite{coe2015, sci2018}. As a consequence of the HK theorem all the observables are a density functional and therefore may in principle be obtained via DFT calculations. The theorem does not provide however any hint of how to obtain both, density of the system and density functional of any desired property. 

The Kohn-Sham (KS) scheme \cite{Kohn1965} is a clever iterative mapping commonly used to obtain the density of an interacting system. It maps the many-body interacting system into a ficticious non-interacting one via the so-called KS potential $v_{KS}$. Here $v_{KS}$ is constructed to reproduce the density $ n $\textbf{(r)} of the interacting system. Since $v_{KS}$ depends on the exchange-correlation energy and the latter is a density functional typically unknown exactly, in practical calculations one needs to make use of approximations to obtain the particle density. Hence, the performance of the DFT results depends on the approximations and functionals used.                             

Therefore there is a very active community in DFT dedicated to the development and the optimization of density functionals for practical DFT calculations. Less exploited tools in  this research line are however the artificial neural networks (ANN) \cite{Bishop1995}. ANN are computational models that refer to the human brain functioning and are capable of calculating non-linear mathematical functions. It resembles the biological brain: it is able to be trained with a dataset and afterwards generalize the learning to other data of the same problem, being a type of machine learning used to estimate parameters and to recognize patterns. 

Among the advantages of ANN one can mention the precision of the results, the easy implementation, the fast calculations and the ability of generalizing from examples \cite{Nafey2009}. Once an ANN has been trained and validated satisfactorily, it is able to predict new outputs from different input data (in the same domain, but different from those used in the training). ANN are used in many areas including neurocomputation, chemical engineering, industrial applications, medicine, chemistry and physics \cite{ortega2017,erica2015,Ramil2018,Walczak2018,Balabin2009, burke, gibbon, fletcher}.

Here we apply ANN's concepts to estimate the ground-state energy of fermionic interacting particles in homogeneous chains as described by the one-dimensional Hubbard model. We designed an ANN {functional} which is able to recover the numerically exact energy of the Hubbard by less than $0.15\%$ for all the regimes of filling factors, $0\leq n\leq 1$, all the regimes of magnetization, $0\leq m \leq n$, and a vast regime of interactions, $0\leq U\leq 10t$, including strongly correlated materials, with typical interaction $U\sim 6t$, where $t$ is the hopping term. Our ANN {functional} was also proven to be superior to current parameterizations, used as input in local and non-local approximations for DFT methods. {Additionally, we have used the homogeneous ANN functional as input in DFT calculations for investigating finite, localized impurities and harmonically confined systems. We find that while our ANN calculations are notably more accurate than other similarly simple and fast DFT approaches, it speeds-up considerably the numerical calculations.} Thus the ANN model proposed here could be used as a practical and reliable tool in DFT calculations for such inhomogeneous fermionic systems, {including} superlattices \cite{superlattice1, superlattice2} and disordered chains \cite{disorder1, disorder2}, for which exact calculations are nontrivial.

\section*{Theoretical and computational methods}

	We consider one-dimensional homogeneous chains described by the fermionic Hubbard model \cite{Hubbard1963},

\begin{eqnarray} \label{hhubbard}
	\hat H &= &  - t\sum\limits_{<ij>\sigma }{(\hat c_{i\sigma }^\dag \hat c_{j\sigma } + H.c.)} + U\sum\limits_i {{\hat n}_{i \uparrow }}{{\hat n}_{i \downarrow }},
\end{eqnarray}
where $U$ is the on-site interaction and $t$ is the nearest-neighbor hopping term, while $\hat c_{i\sigma }^\dag$ ($\hat c_{i\sigma }$) is the creation (annihilation) operator of fermions at site $i$ with spin $\sigma=\uparrow, \downarrow$, and ${\hat n_{i\sigma }} = \hat c_{i\sigma }^\dag \hat c_{i\sigma }$ the number operator. The average {charge} density or filling factor is given by $n = N/L= \left\langle {{{\hat n}_ \uparrow }} \right\rangle  + \left\langle {{{\hat n}_ \downarrow }} \right\rangle$, while the magnetization {or spin density} is $m= \left\langle {{{\hat n}_ \uparrow }} \right\rangle  -\left\langle {{{\hat n}_ \downarrow }} \right\rangle$, where $N = {N_ \uparrow } + {N_ \downarrow }$ is the total number of particles and $L$ the chain size.

This is the simplest model to describe itinerant and interacting electrons in a chain, such as in solids and in nanostructures \cite{coe2010, coe2011, coeEPL}, or cold atoms in an optical lattice \cite{thj,  vvf2008, vvf2012}. Despite its simplicity the Hubbard model describes important phenomena \cite{vivialdo}, such as the Mott transition from metal to insulator (for $U>0$, at $n=1$) \cite{vvf2006}, the transition from Bardeen-Cooper-Schrieffer superfluid to Bose-Einstein condensate (for $U<0$, at $m=0$) \cite{vvf2006} and exotic superfluidity (for $U<0$ and $m\neq0$) \cite{vvf2012, vvf2017, picoli}.

Nevertheless the exact analytical solution of the model is in general unknown $-$ except for some specific parameters \cite{Franca2012} $-$ thus one has to perform numerical calculations. For spatially homogeneous chains, $L=\infty$, one can obtain an exact fully numerical (FN) solution by solving the Lieb-Wu integrals \cite{lieb}. For finite chains, using exact diagonalization we are limited to small chains, $L\lesssim 15$, while with density-matrix renormalization group (DMRG) \cite{dmrg}, which is almost exact, one can solve chains of $L\lesssim 200$ sites. DMRG are however computationally expensive calculations: a single calculation might take several hours in a high-performance computing cluster.  

Thus DFT appears as a powerful tool to solve larger chains at a lower computational cost: a typical DFT calculation for the Hubbard model takes no longer than 1 minute. DFT may also be applied to solve inhomogeneous chains, i.e., in the presence not only of boundaries, but also impurities, interfaces, superlattices, harmonic confinement and disorder. This is possible thanks to $-$ local or non-local $-$ approximations to the exchange-correlation functional, which require, as an input, a {functional} for the per-site ground-state energy of the homogeneous system.

{In particular, the local-spin-density approximation (LSDA) for the ground-state energy is given by}

\begin{equation}
{
e_0^{inh}\approx e_0^{LSDA}=\frac{1}{L} \left.\sum_{i=1}^L{ e^{hom}_0[n,m,U]}\right|_{^{n \rightarrow n_i}_{m \rightarrow m_i}}}\label{lsda}
\end{equation}
{where $e_0^{hom}$ is the homogeneous functional for the ground-state energy density and $(n_i, m_i)$ are local densities at site $i$ of the inhomogeneous system. Thus Eq.(\ref{lsda}) approximates the energy of inhomogeneous chains by evaluating the energy of the homogeneous system with the site-by-site replacement $n\rightarrow n_i$ and $m\rightarrow m_i$. In practical DFT, using the self-consistent KS scheme\cite{Kohn1965}, it implies the evaluation of partial derivatives of the energy with respect to $n_i$ and $m_i$ (to obtain $v_{KS}$) in each of the iterations of the self-consistency cycle. Therefore, in order to solve inhomogeneous Hubbard chains within DFT-LSDA, the first step is to obtain a simple and reliable functional $e_0^{hom}[n,m,U]$ for the homogeneous limit.}

{Although the fully numerical (FN) solutions of the Lieb-Wu integrals\cite{lieb} (exact only for infinite chains) are certainly a trusty functional for the DFT-LSDA treatment of inhomogeneous systems, they are computationally expensive. Besides, the integrals are not direct functions of (n,m) and hence we do not have total freedom to choose specific values to be used site by site in each iteration of the KS cycles. Therefore, for each interaction $U$, one has to produce a huge amount of FN data to cover all the regimes of $n_i$ and $m_i$ in order to obtain the $v_{KS}$ numerical derivatives with high precision.} 

{So one possibility to overcome this numerical difficulty is then to adopt a reliable analytical expression for the homogeneous functional. Actually it has been done:} the {analytical} FVC functional \cite{Franca2012}{, which} extended previous expressions \cite{Lima2003} by incorporating the magnetization{, has been currently used as a practical tool for DFT-LSDA calculations}\cite{disorder1, xian2012, paula, JPCM17, carsten}. The FVC parametrization properly recovers all the analytical known limiting cases (for $U=0$, $U=\infty$ and for the combined set $n=1$ and $m=0$), correctly predicts a positive Mott gap at $n=1$ for any $U > 0$, and is also more precise in general than the previous expressions \cite{Lima2003}. While for most of the parameters the FVC {functional} presents relatively low deviations from numerically exact results ($\sim 2\%$), it shows however inopportune larger deviations for the weakly interacting regime. {Thus, the ANN functional proposed here is a simpler and faster numerical approach to inhomogeneous systems within DFT calculations, with superior accuracy than FVC and computationally cheaper than FN and DMRG.}

\section*{Results and discussion}

The training of our neural model was performed with the Levenberg-Marquartd algorithm \cite{Hagan1994}. The process involves the random split of the exact data in three sets: 70\% for the training itself, 15\% for the validation and the remaining 15\% for testing the ANN performance. The input parameters of our ANN model are the variables density $n$, magnetization $m$ and interaction $U$, while the output is the per-site ground-state energy $e^{ANN}_0(n,m,U)$. We consider 20,891 numerically exact Lieb-Wu results within the regimes $0 \leq U \leq 10t$, $0 \leq n \leq 1$ and $0 \leq m \leq n$. Thus the possible topologies of our ANN are of type: 3 input neurons $\{n,m,U\}$ $-$ a certain number of hidden layers $-$ a single output neuron $\{e_0^{ANN}(n,m,U)\}$. Several topologies were studied, e.g., with a single hidden layer $-$ 3-5-1, 3-10-1 and 3-20-1 $-$ and with two hidden layers, 3-12-12-1 and 3-10-10-1.

\begin{figure}[!b]
	\centering
	\includegraphics[width=0.83\linewidth]{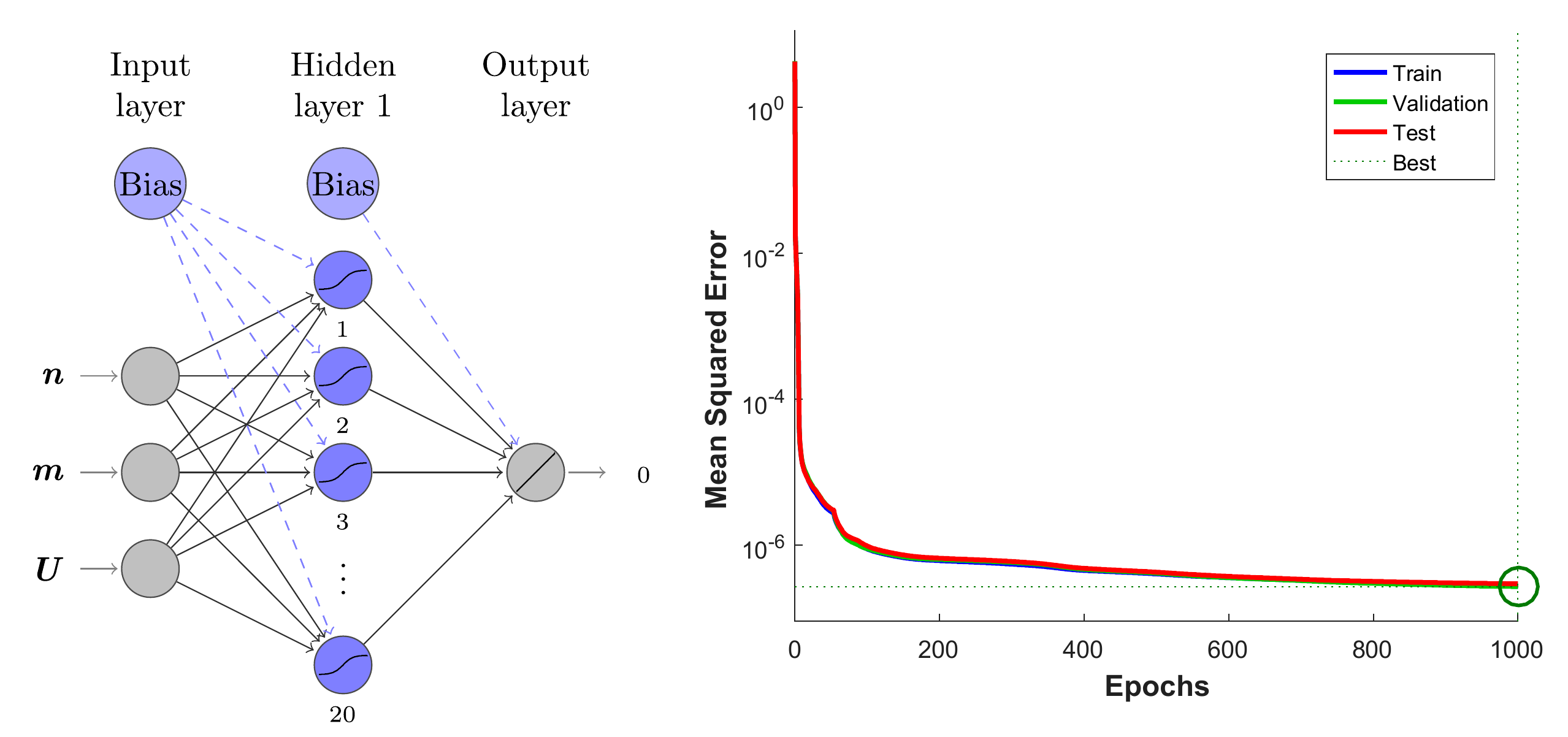}
	\caption{Structure of our ANN model, 3-20-1, and its performance during training.}
	\label{fig:Figure-1}
\end{figure}	

Our target performance was below $10^{-6}$ for the mean squared error, which was obtained after extensive training by the topology 3-20-1 with $1000$ epochs (we choose to stop the training in 1000 epochs to avoid over-fitting by the network, which could lead to a poorer generalization performance), as shown in Figure \ref{fig:Figure-1}. This architecture contains one hidden layer composed of twenty neurons, with tangent sigmoid transfer function, while the output layer has only one node with a linear transfer function. The trained neural network can be found online free of charge as Supplementary Material (ANN\_to\_Hubbard.mat and Read\_me.pdf files).

\begin{figure}[H]
	\centering
	\includegraphics[width=0.66\linewidth]{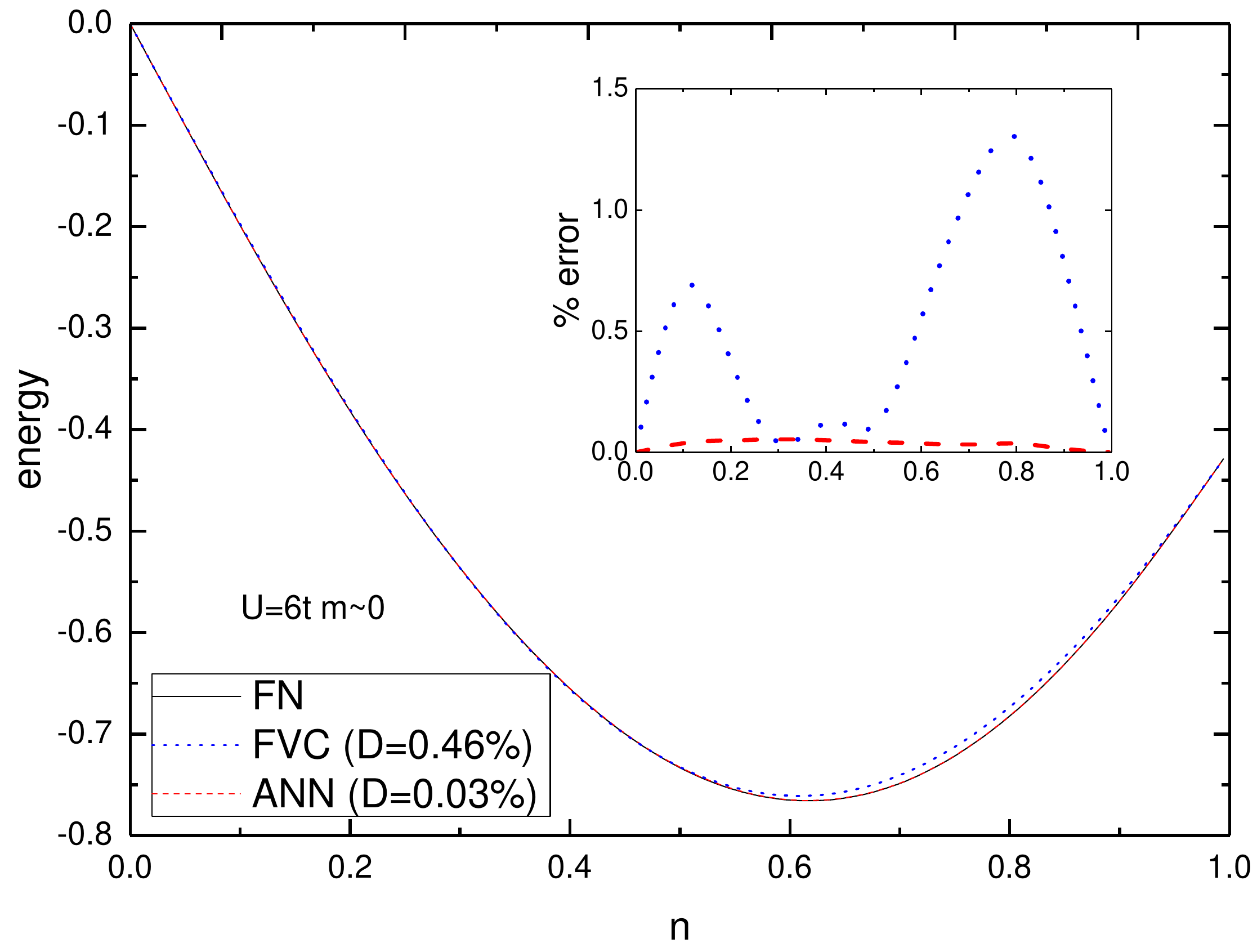}
	\caption{Per-site ground-state energy as a function of the density {for the FVC and ANN functionals in comparison to the numerically exact FN result}. The inset shows the percentage deviations, obtained via $100\times|(e_0^{approx} - e_0^{exact})/e_0^{exact}|$. {In all cases the interaction is $U=6t$, the magnetization is $m\sim0$ and $D$ is the average percentage deviation.}}
	\label{fig:Figure-3}
\end{figure}

Now we compare the perfomance of our ANN model with the FVC parametrization, for all the regimes of parameters. We start by considering non-magnetic systems. In Figure \ref{fig:Figure-3} we show the energy as a function of the density for a strongly correlated case, with $U=6t$. While the overall behavior is well described by both, ANN and FVC, we find that the ANN deviations are below $\sim 0.1\%$ (see inset), smaller than the FVC ones, which in some cases are $10$ times bigger. This higher performance of ANN over FVC for strongly interacting systems is surprising, since at this regime FVC is recognized as sufficiently precise \cite{Franca2012}.

\begin{figure}[H]
	\centering
	\includegraphics[width=0.66\linewidth]{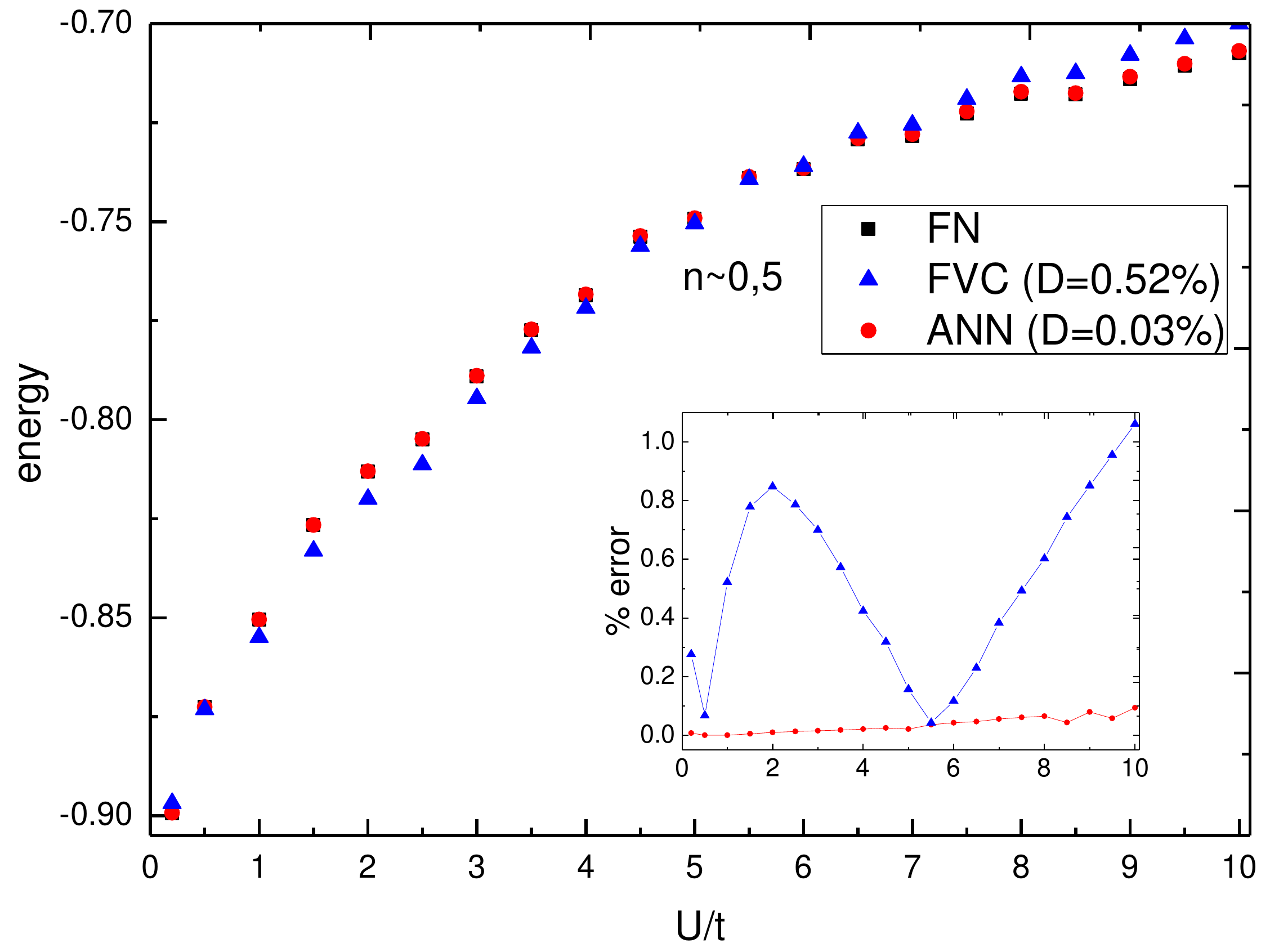}
	\caption{Per-site ground-state energy as a function of the on-site interaction $U$ (in units of $t$), {for the FVC and ANN functionals in comparison to the numerically exact FN result. Here $n\sim0.5$ and magnetization $m\sim0$}. The inset shows the percentage deviations, obtained via $100\times|(e_0^{approx} - e_0^{exact})/e_0^{exact}|${, while $D$ is the average percentage deviation.}}
	\label{fig:Figure-4}
\end{figure}	

Figure \ref{fig:Figure-4} presents the energy as a function of the interaction for non-magnetic chains. It reveals that the excellent performance of our ANN model is not restricted to the strongly correlated systems: within $0\leq U\leq 10t$ and $m\sim 0$ the ANN deviates by less than $\sim0.08\%$ for all the interactions, while FVC deviates on average $\sim 0.5\%$.

\begin{figure}[H]
	\centering
	\includegraphics[width=0.66\linewidth]{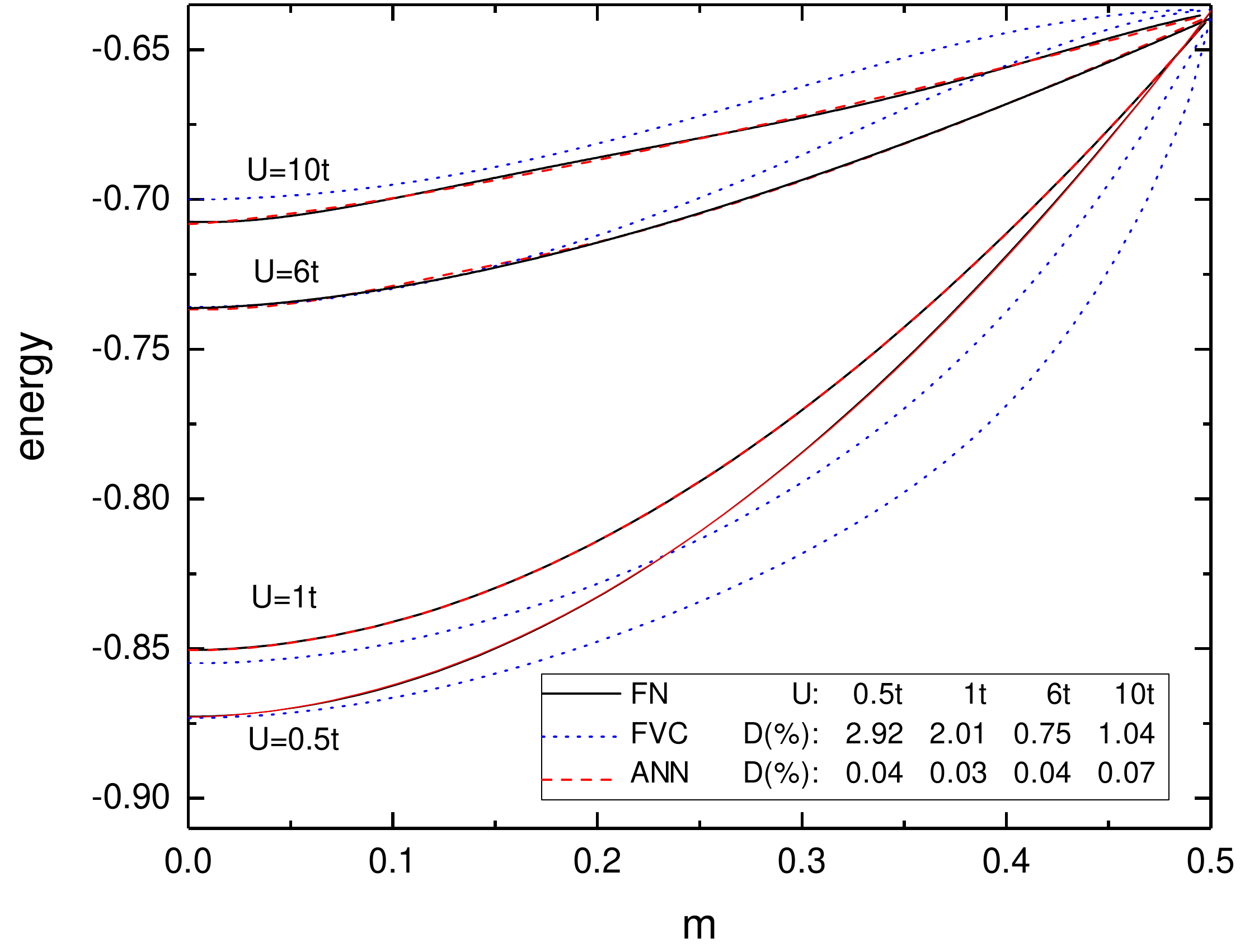}
	\caption{{Per-site ground-state} energy as a function of the magnetization \textit{m}, {for the FVC and ANN functionals in comparison to the numerically exact FN result. Here $n\sim0.5$ and $D$ is the average percentage deviation.}
} 	\label{fig:Figure-5}
\end{figure}

{We also} compare the performance of ANN and FVC for magnetic systems. {Figure \ref{fig:Figure-5} presents} the energy as a function of the magnetization for several interactions. We find that while FVC in some cases fails to recover even the qualitative behavior of the energy, our ANN model properly describes the exact trend. {Quantitatively, while} FVC deviations reach up to $\sim 7\%$ {(for $U=0.5t$)}, the ANN model remains reliable, with deviations inferior to $0.15\%$.

{Finally, we illustrate the use of our ANN functional within DFT-LSDA calculations by applying it to inhomogeneous \textit{i)} unconfined finite chains, \textit{ii)} localized impurity systems ($V_i=V\delta_{i,i_V}$)  and \textit{iii)} harmonically confined chains ($V_i=k(i-i_0)^2$). The later inhomogeneous potential is particularly interesting to simulate state-of-the-art experiments with cold fermionic atoms in optical lattices. We compare the performance of DFT-LSDA $-$ with each functional ANN, FVC and FN $-$ to DMRG calculations.}

\begin{figure}[H]
	\centering
	\hspace{-1.5cm}
	\includegraphics[width=0.98\linewidth]{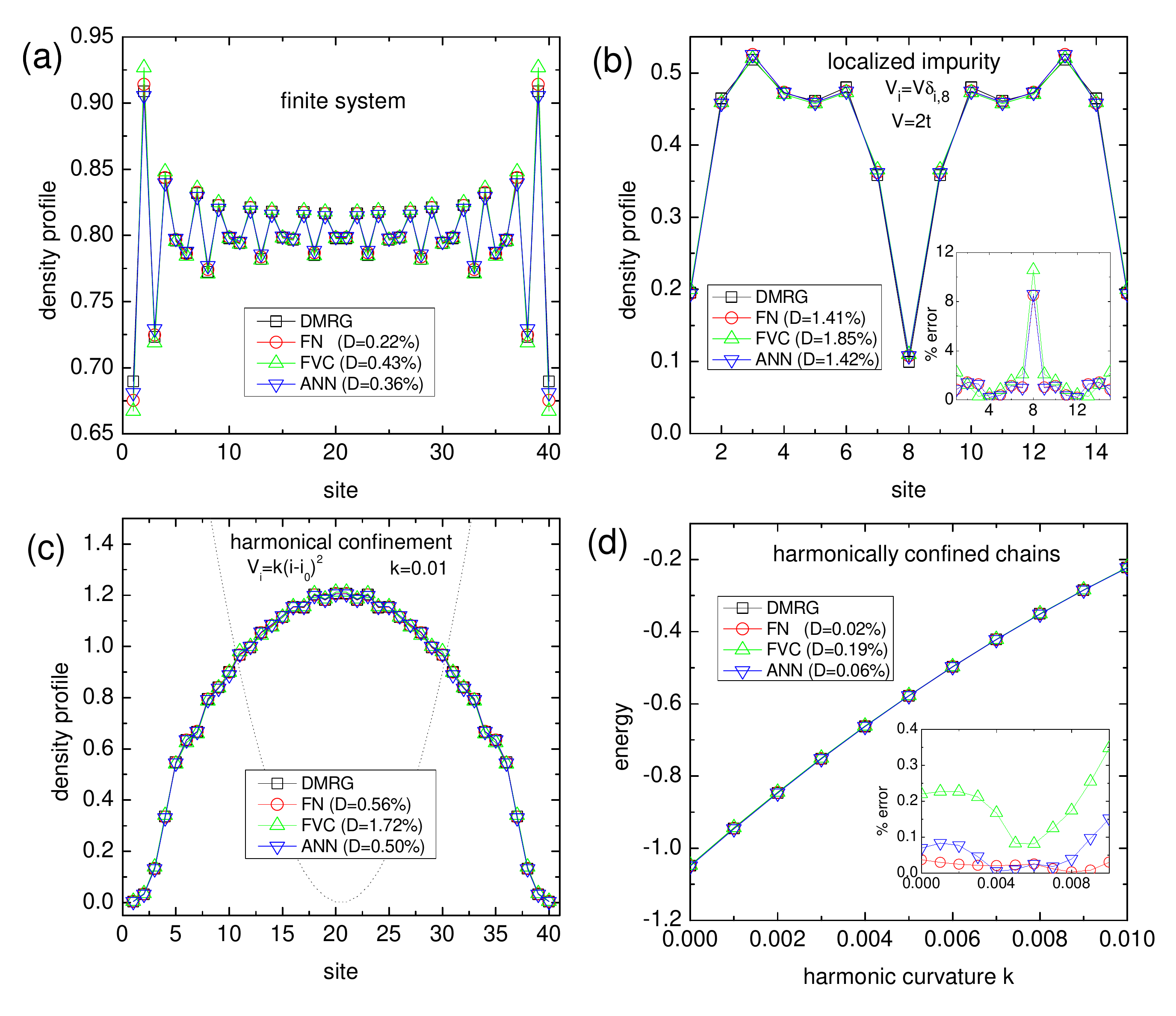}\hspace{-1.2cm}
		\caption{{Density profile for (a) a finite chain of size $L=40$ and $n=0.8$, (b) a localized impurity system, with $L=15$, impurity strength $V=2t$ and $n=0.4$, and (c) a harmonically confined chain with $L=40$, harmonic curvature $k=0.01$ and $n=0.8$, while (d) shows the per-site ground-state energy as a function of the harmonic confinement curvature for chains of size $L=40$ and $n=0.8$. In all cases we adopted open boundary conditions, $U=1t$ and $m\sim0$. The results are either exact DMRG or approximated DFT-LSDA, with one of the functionals: FN, FVC or ANN. The percentage deviations are obtained via $100\times|(e_0^{approx} - e_0^{exact})/e_0^{exact}|$ and $D$ is the average percentage deviation.}}
	\label{fig:Figure-6}
\end{figure}

{Figure \ref{fig:Figure-6} summarizes the essential features of our analysis to the several inhomogeneous test cases. For finite unconfined chains, Fig.~\ref{fig:Figure-6}a, the typical Friedel-like oscillations \cite{Friedel} on the local densities are in general well reproduced by all the three functionals. The average percentage deviations in this case are similar, slightly larger for FVC, because the chain is relatively large, $L=40$. For smaller chains and/or with inhomogeneous external potentials $V_i$, as in Fig.~\ref{fig:Figure-6}b and Fig.~\ref{fig:Figure-6}c, we see that the FVC is clearly the less accurate functional. Fig.~\ref{fig:Figure-6}d, for the harmonically confined case, shows that the deviations on the ground-state energy follow similar trend: although the deviation oscillates with the harmonic strength $k$ for the approaches, the FVC deviations are systematically larger than the FN and ANN ones. Thus, our results demonstrate that the ANN approach has a very good balance between precision and computational cost: it is faster than FN and DMRG and significantly more precise than FVC.}

\section*{Conclusions}
In summary, we have developed a simple, fast and reliable artificial neural network {functional} for the ground-state energy of the homogeneous fermionic Hubbard chains. {We have exemplified the powerfulness of the ANN functional into DFT methods by exploring finite, localized impurities and harmonically confined systems.} The results show that {our ANN approach} not only properly captures the qualitative behavior of energy {and density profiles for all the inhomogeneous systems, but also leads to} more accurate results than previous parameterizations. {This excellent performance is similar to other numerical solutions (FN and DMRG), but at a fraction of the computational cost}. Therefore our ANN {functional is a practical tool to be implemented in DFT calculations}, such as in local and non-local density approximations, for inhomogeneous Hubbard chains. Our approach is particularly interesting to investigate disordered systems, where many realizations are needed for each randomly localized impurities scenario, what makes analyses via exact methods computationally prohibitive.

\section*{Data Availability}
The datasets generated during and/or analysed during the current study are available from the corresponding author on reasonable request.

\begin{thebibliography}{00}
\bibitem{tobi}
Bruenner, T., Runge, E., Buchleitner, A. \& Fran\c{c}a, V. V. Entanglement enhancement in spatially inhomogeneous many-body systems. \textit{Phys. Rev. A} \textbf{87}, 3231 (2013).

\bibitem{PhysRev136B864}
Hohenberg, P. \& Kohn, W. Inhomogeneous Electron Gas. \textit{Phys. Rev.} \textbf{136}, B864-B871 (1964).

\bibitem{Kohn1999}
Kohn, W. Nobel Lecture: Electronic structure of matter - wave functions and density functionals.\textit{ Rev. Mod. Phys.} \textbf{71}, 1253-1266 (1999).

\bibitem{Capelle2006}
Capelle, K. A Bird's-Eye View of Density-Functional Theory.\textit{ Braz. J. Phys.} \textbf{36}, 1318-1343 (2006). 

\bibitem{coe2015}
Coe, J. P., D'Amico, I. \& Fran\c{c}a, V. V. Uniqueness of density-to-potential mapping for fermionic lattice systems. \textit{EPL Europhysics Lett.} \textbf{110}, 63001 (2015).

\bibitem{sci2018}
Fran\c{c}a, V. V., Coe, J. P., \& D'Amico, I. Testing density-functional
approximations on a lattice and the applicability of the related Hohenberg-Kohn-like theorem.  \textit{Sci. Rep.} \textbf{8}, 664 (2018).

\bibitem{Kohn1965}
Kohn, W. \& Sham, L. J. Self-consistent equations including exchange and correlation effects. \textit{Phys. Rev.} \textbf{140}, A1133-A1138 (1965).

\bibitem{Bishop1995}
Bishop, C. M. \textit{Neural networks for pattern recognition} (Clarendon Press, 1995).

\bibitem{Nafey2009}
Nafey, A. S. Neural network based correlation for critical heat flux in steam-water flows in pipes. \textit{Int. J. Therm. Sci.} \textbf{48}, 2264-2270 (2009).

\bibitem{ortega2017}
Ortega-Zamorano, F., Jerez, J. M., Ju\'{a}rez, G. E. \& Franco, L. FPGA Implementation of Neurocomputational Models: Comparison Between Standard Back-Propagation and C-Mantec Constructive Algorithm. \textit{Neural Process. Lett.} \textbf{46}, 1-16 (2017).

\bibitem{erica2015}
Filletti, E. R., Silva, J. M. \& Ferreira, V. G. Predicting of the fibrous filters efficiency for the removal particles from gas stream by artificial neural network. \textit{Adv. Chem. Eng. Sci.} \textbf{05}, 317-327 (2015).

\bibitem{Ramil2018}
Ramil, A., Lopez, A., Pozo-Antonio, J. \& Rivas, T. A computer vision system for identification of granite-forming minerals based on rgb data and artificial neural networks. \textit{Measurement} \textbf{117}, 90-95 (2018).

\bibitem{Walczak2018}
Walczak, S. \& Velanovich, V. Improving prognosis and reducing decision regret for pancreatic cancer treatment using artificial neural networks.\textit{ Decis. Support. Syst.} \textbf{106}, 110-118 (2018).

\bibitem{Balabin2009}
Balabin, R. M. \& Lomakina, E. I. Neural network approach to quantum-chemistry data: Accurate prediction of density functional theory energies. \textit{J. Chem. Phys.} \textbf{131} (2009).

\bibitem{burke}
Brockherde, F. et al. Bypassing the Kohn-Sham equations with machine learning. \textit{Nat. Commun.} \textbf{8}, 872 (2017).

\bibitem{gibbon}
McGibbon, R. T. \& Pande, V. S. Learning Kinetic Distance Metrics for Markov State Models of Protein Conformational Dynamics. \textit{J. Chem. Theory Comput.} \textbf{9}, 2900-2906 (2013).

\bibitem{fletcher}
Fletcher, T. L., Davie, S. J. \& Popelier, P. L. A. Prediction of Intramolecular Polarization of Aromatic Amino Acids Using Kriging Machine Learning. \textit{J. Chem. Theory Comput.} \textbf{10}, 3708-3719 (2014).

\bibitem{superlattice1} 
Zhang, L.-L., Huang, J., Duan, C.-B. \& Wang, W.-Z. Incommensurate charge density wave modulated by period of Hubbard superlattices. \textit{Mod. Phys. Lett. B} \textbf{29}, 1550208 (2015).

\bibitem{superlattice2}
Wei-Zhong, Z. L.-L., Jin, H., Cheng-Bo, D. \& Wang, W.-Z. Structure-dependent metal?insulator transition in one-dimensional Hubbard superlattice. \textit{Chin. Phys. B} \textbf{24}, 77101 (2015).

\bibitem{disorder1}
{Vettchinkina, V., Kartsev, A., Karlsson, D., \& Verdozzi, C. Interacting fermions in one-dimensional disordered lattices: Exploring localization and transport properties with lattice density-functional theories. \textit{Phys. Rev. B} \textbf{87}, 115117 (2013).}

\bibitem{disorder2}
Fran\c{c}a, V. V. \& D'Amico, I. Entanglement from density measurements: Analytical density functional for the entanglement of strongly correlated fermions. \textit{Phys. Rev. A} \textbf{83}, 42311 (2011).

\bibitem{Hubbard1963}
Hubbard, J. Electron Correlations in Narrow Energy Bands. \textit{Source: Proc. Royal Soc. London. Ser. A Math. Phys. Sci.} \textbf{276}, 238-257 (1963).

\bibitem{coe2010}
Coe, J. P., Fran\c{c}a, V. V. \& D'Amico, I. Hubbard model as an approximation to the entanglement in nanostructures. \textit{Phys. Rev. A} \textbf{81}, 052321 (2010).

\bibitem{coe2011}
Coe, J. P., Fran\c{c}a, V. V. \& D'Amico, I. Approximation of the entanglement in quantum dot chains using Hubbard models. \textit{J. Physics: Conf. Ser.} \textbf{286}, 12048 (2011).

\bibitem{coeEPL}
{Coe, J. P., Fran\c{c}a, V. V. \& D'Amico, Feasibility of approximating spatial and local entanglement in long-range interacting systems using the extended Hubbard model. \textit{EPL} \textbf{93}, 10001 (2011).}

\bibitem{thj}
Johnson, T. et al. Hubbard Model for Atomic Impurities Bound by the Vortex Lattice of a Rotating Bose-Einstein Condensate. \textit{Phys. Rev. Lett.} \textbf{116}, 240402 (2016).

\bibitem{vvf2008}
Fran\c{c}a, V. V. \& Capelle, K. Entanglement in spatially inhomogeneous many-fermion systems. \textit{Phys. Rev. Lett.} \textbf{100}, 1-4 (2008).

\bibitem{vvf2012}
Fran\c{c}a, V. V., H\"{o}rndlein, D. \& Buchleitner, A. Fulde-Ferrell-Larkin-Ovchinnikov critical polarization in one-dimensional fermionic optical lattices. \textit{Phys. Rev. A} \textbf{86}, 033622 (2012).

\bibitem{vivialdo}
Capelle, K. \& Campo, V. L. Density functionals and model Hamiltonians: Pillars of many-particle physics. \textit{Phys. Reports} \textbf{528}, 91-159 (2013).

\bibitem{vvf2006}
Fran\c{c}a, V. V. \& Capelle, K. Entanglement of strongly interacting low-dimensional fermions in metallic, superfluid, and antiferromagnetic insulating systems. \textit{Phys. Rev. A} \textbf{74}, 42325 (2006).

\bibitem{vvf2017}
Fran\c{c}a, V. V. Entanglement and exotic superfluidity in spin-imbalanced lattices. \textit{Phys. A: Stat. Mech. its Appl.} \textbf{475}, 82-87 (2017).

\bibitem{picoli}
De Picoli, T., D'Amico, I. \& Fran\c{c}a, V. V. Metric-space approach for distinguishing quantum phase transitions in spin-imbalanced systems. {\textit{Braz. J. Phys.} \textbf{48}, 472 (2018).} 

\bibitem{Franca2012}
Fran\c{c}a, V. V., Vieira, D. \& Capelle, K. Simple parameterization for the ground-state energy of the infinite Hubbard chain incorporating Mott physics, spin-dependent phenomena and spatial inhomogeneity. \textit{New J. Phys.} \textbf{14} (2012). 

\bibitem{lieb}
Lieb, E. H. \& Wu, F. Y. Absence of Mott Transition in an Exact Solution of the Short-Range, One-Band Model in One Dimension. \textit{Phys. Rev. Lett.} \textbf{20}, 1445-1448 (1968).

\bibitem{dmrg}
Schollw\"{o}ck, U. The density-matrix renormalization group. \textit{Rev. Mod. Phys.} \textbf{77}, 259-315 (2005).

\bibitem{Lima2003}
Lima, N. A., Silva, M. F., Oliveira, L. N. \& Capelle, K. Density functionals not based on the electron gas: Local-density approximation for a luttinger liquid. \textit{Phys. Rev. Lett.} \textbf{90}, 146402 (2003).

\bibitem{xian2012}
Xianlong, G., Chen, A.-H., Tokatly, I. V., \& Kurth, S. Lattice density functional theory at finite temperature with
strongly density-dependent exchange-correlation potentials. \textit{Phys. Rev. B} \textbf{86}, 235139 (2012).

\bibitem{paula}Mori-Sanchez, P., \& Cohen, A. J. The derivative discontinuity of the exchange?correlation functional, \textit{Phys. Chem. Chem. Phys.} \textbf{16}, 14378 (2014).

\bibitem{JPCM17} Mitxelena, I., Piris, M., \& Rodriguez-Mayorga, M. On the performance of natural orbital functional approximations in the Hubbard model. \textit{J. Phys.: Condens. Matter} \textbf{29}, 425602 (2017).

\bibitem{carsten} Ulrich, C. A. Density-functional theory for systems with noncollinear spin: Orbital-dependent exchange-correlation functionals and their application to the Hubbard dimer, \textit{Phys. Rev. B} \textbf{98}, 035140 (2018).

\bibitem{Hagan1994}
Hagan, M. T. \& Menhaj, M. B. Training Feedforward Networks with the Marquardt Algorithm. \textit{IEEE Transactions on Neural Networks} \textbf{5}, 989-993 (1994).

\bibitem{Friedel} {Vieira, D., Freire, H. J. P., Campo Jr., V. L. \& Capelle, K. Friedel oscillations in one-dimensional metals: From Luttinger's theorem to the Luttinger liquid. \textit{J. Magn. Magn. Mater.} \textbf{320}. E421-E424 (2008).}

\end{thebibliography}

\section*{Acknowledgements}

VVF acknowledges FAPESP (Grant: 2013/15982-3), VVF and ERF acknowledge CNPq (Grant: 448220/2014-8). {This research was supported by resources supplied by the Center for Scientific Computing (NCC/GridUNESP) of the S\~{a}o Paulo State University (UNESP).}

\section*{Contributions}

All authors conceived the idea, C.A.C. performed the calculations. All authors discussed the results and contributed to the writing of the manuscript.

\section*{Additional information}


\subsection*{Competing financial interests}
The authors declare that they have no competing interests.

\end{document}